\documentclass[oldversion]{aa} 
\usepackage{graphicx}
\usepackage{aalongtable}
\usepackage{txfonts}
\usepackage{rotating}
\usepackage{lscape}
\newcommand{\gsim}{\;\lower.6ex\hbox{$\sim$}\kern-7.75pt\raise.65ex\hbox{$>$}\;}
\newcommand{\lsim}{\;\lower.6ex\hbox{$\sim$}\kern-7.75pt\raise.65ex\hbox{$<$}\;}

\begin{document}
\title{The fraction of first and second generation stars in globular
clusters. I The case of NGC~6752\thanks{Table 2 is only available in electronic form at the CDS via anonymous
   ftp to {\tt cdsarc.u-strasbg.fr} (130.79.128.5) or via
   {\tt http://cdsweb.u-strasbg.fr/cgi-bin/qcat?J/A+A/???/???}} 
 }

\author{Eugenio Carretta
}

\authorrunning{E. Carretta}
\titlerunning{Fraction of first and second generation stars}

\offprints{E. Carretta, eugenio.carretta@oabo.inaf.it}

\institute{
INAF-Osservatorio Astronomico di Bologna, Via Ranzani 1, I-40127
 Bologna, Italy
}

\date{}

\abstract{We present a new method to estimate the fraction of stars with
chemical composition of first and second(s) generation(s) currently hosted in
Galactic globular clusters (GCs). We compare cluster and field stars of similar
metallicity in the [Fe/H]-[Na/H] plane. Since the phenomenon of multiple
populations is only restricted to the cluster environment, the number of GC
stars whose location coincides with that of field stars provides the fraction of
first generation stars in that cluster. By exclusion, the fraction of second
generation stars is derived. We assembled a dataset of 1891 field stars of the
thin disk, thick disk, and halo of the Milky Way in the metallicity range $-3.15
\leq$ [Fe/H] $\leq +0.48$ dex and with Na abundance from high resolution
spectra. They are mostly dwarfs, but include also giants. Considering only the
range in metallicity spanned by most GCs extensively studied for the Na-O
anticorrelation ($-2.36 \leq$ [Fe/H] $\leq -0.33$ dex), we have 804 stars. The
total sample is homogeneized by offsets in [Fe/H] and [Na/H] with respect to a
reference sample using the same line list and NLTE correction for Na adopted in
a recent extensive survey of GC stars. This fully accounts for offsets among
analyses due to different temperature scales, line lists, adopted (or neglected)
corrections for departures from LTE. We illustrate our method estimating the
fraction of first and second generation stars in the well studied GC NGC~6752.
As a by-product, the comparison of [Na/H] values in GC and field stars suggests
that at least two classes of old stellar systems probably contributed to the
halo assembly: one group with characteristics similar to the currently existing
GCs, and the other more similar to the present-day dwarf satellite galaxies.
}

\maketitle

\section{Introduction}

Once upon a time, the Galactic globular clusters (GCs) were considered the best
example in nature of simple stellar populations (SSP, see e.g. the review by
Renzini and Fusi Pecci 1988). The pioneering studies by Cohen (1978) and
Peterson (1980) however showed that supposedly coeval stars in GCs were not of
the same initial chemical composition, which is a requisite, by definition, of
SSPs. In particular, Cohen (1978) concluded that ``the primordial gas in
M3 was not chemically homogeneous during the time interval when star formation
of the currently observed M3 red giants took place." 

In both studies a scatter of Na abundances in cluster stars was noted, and Na is
still one of the best tracers used to characterize the peculiar chemistry of
stars in GCs. Different amounts of Na excess, anticorrelated with variyng
degrees of O depletion, were found in all GCs studied by the Lick-Texas group
(see Kraft 1994 for a review). The discovery of the Na-O anticorrelation,
coupled to the C-N anticorrelation already widely studied among cluster giants
(e.g. Smith 1987), led to identify the nucleosynthesis source responsible for
the observed changes in the network of proton-capture reactions in H-burning at
high temperatures. The NeNa cycle was expected to operate, producing Na, in the
same region where the ON part of the CNO cycle is fully active (Denisenkov and
Denisenkova 1989, Langer et al. 1993), destroying O. The abundance variations
were found only restricted to the dense environment of globular clusters, since
Gratton et al. (2000) demonstrated that Na, O abundances are not modified in
field stars from the typical levels established by supernovae (SNe)
nucleosynthesis only. 

However, at the time it was unclear where the hot H-burning was located, whether
in the stars currently observed in GCs or outside. The finding of the Na-O
anticorrelation (as well as star-to-star variations in Mg anticorrelated to
those in Al) among unevolved stars in NGC~6752 (Gratton et al. 2001) 
unambigously answered this question. These scarcely evolved stars do not reach
temperatures high enough to efficiently produce Na or  Al in their interiors;
moreover, their convective envelope interest a negligible fraction of their
mass. The milestone established by Gratton et al. (2001) was to show that the
chemical pattern observed in dwarfs currently in GCs $must$ have been
necessarily imprinted in the gas from which they formed by the most massive
stars of a previous stellar generation. Observing Na-O variations in cluster
stars means that we are in presence of multiple stellar generations, no more and
no less.

The new paradigma for GCs as examples of multiple stellar populations is 
currently well assessed, and much
work was made in recent years (see the exaustive reviews by Gratton et al. 2004,
2012): chemical tagging of different generations, identification of multiple
photometric sequences in the color-magnitude diagrams (CMDs), impact of modified
chemical composition on photometric bands (e.g. Carretta et al. 2011, Sbordone
et al. 2011, Milone et al. 2012, Cassisi et al. 2013). However, much still
remains to be done, in particular to better quantify the properties of different
stellar generations such as the ratio between the fraction of first and 
second(s) generation(s) stars currently composing the cluster populations.
This ratio is a fundamental parameter in several key issues related to the
formation of GCs, their dynamical evolution and spatial mixing (Vesperini et al.
2013), and their connection to the formation of the Galactic halo (Carretta et
al. 2010,  Vesperini et al. 2010, Martell et al. 2011).

The first quantitative estimate was possible thanks to the extensive FLAMES
survey of GCs (Carretta et al. 2006, Carretta et al. 2009a,b, and
ongoing follow up). The homogeneous dataset of Na, O abundances for more than
1500 giant stars in about 20 GCs allowed Carretta et 
al. (2009a) to show that only one third of stars currently observed in GCs has a
chemistry of first generation, primordial stars, the bulk ($\sim 70\%$) belongs
to the second generation. Stars were assigned to the primordial (P) component if
their O, and Na, content was similar to that of field stars of the same
metallicity [Fe/H]\footnote{We adopt the usual spectroscopic notation, $i.e.$ 
[X]= log(X)$_{\rm star} -$ log(X)$_\odot$ for any abundance quantity X, and 
log $\epsilon$(X) = log (N$_{\rm X}$/N$_{\rm H}$) + 12.0 for absolute number
density abundances.}. Stars deviating from the high-O, low-Na locus were
considered second generation stars, of an intermediate (I) or extreme (E)
component according to their [O/Na] ratio. 

Recently, Milone et al. (2013) criticized as arbitrary the criteria used in
Carretta et al. (2009a), because in the  case of NGC~6752 the three discrete
groups identified in Carretta et al. (2011, 2012) did not match the
sub-divisions found from the abundances by Yong et al. (2003, 2005). However,
the separation between first and second generation in Carretta et al. (2009a) is
not arbitrary; the P component includes all stars with [Na/Fe] ratios in the
range between [Na/Fe]$_{\rm min}$ and [Na/Fe]$_{\rm min}+0.3$, where 0.3 dex
corresponds to $\sim 4$ times the star-to-star error on [Na/Fe] in each cluster,
hence this group includes all stars with Na abundances similar to that of field
stars\footnote{The separation between the I and E fraction in second generation
stars, at [O/Na]=-0.9 dex, is somewhat arbitrary, but this value corresponds to
the minimum or sudden drop in the distribution function of [O/Na] for GCs with a
long tail of very O-poor stars (NGC~2808, NGC~3201, NGC~5904).}. The method used
in Carretta et al. (2009a) may suffer two possible shortcomings. First, it is
based only on stars on the Na-O anticorrelation in each cluster, i.e. with
measured abundances for $both$ O and Na, even if the separation first/second
generation is made using only Na abundances. Second, the minimum ratio
[Na/Fe]$_{\rm min}$ is estimated by eye, assuming that the stars with lowest Na
abundances along the anticorrelation (excluding possibly a few outliers) have
the typical composition of normal halo stars, with no direct comparison with
field stars.

Hence, in the present study we present a simple variation of that method, to
overcome its possible uncertainties. We propose to compare in the [Na/H] vs
[Fe/H] plane the location of Galactic field stars, fixing the level from SNe
nucleosynthesis only, to the position of giants in GCs. The excess of Na will
clearly select second generation stars with modified composition.
This approach does not rely directly on the Na-O anticorrelation, hence it may
make use of all stars with measured Na abundances in a GC, usually a larger
number than those with O abundances, more difficult to measure (O is depleted, 
whereas Na is enhanced in proton-capture reactions).

In the present paper we will present the assembling, over the metallicity range
spanned by Galactic GCs, of the comparison sample of field stars (Section 2), 
using several literature studies, all shifted on a system defined by our
reference sample (Gratton et al. 2003). The outliers in the [Na/H]-[Fe/H]
distribution are discussed in Section 3, and the application of the present
method is tested in Section 4 using NGC~6752, one of the closest and best
studied GC, as a template. As a by-product of this comparison, some final
considerations on the candidate building blocks that may have contributed to the
formation of the Galactic halo are given in Section 4.

The abundances of Na in the reference sample by Gratton et al. (2003) are 
obtained with the same line lists, and include the same prescriptions for NLTE
corrections to Na values, used in the FLAMES survey of GCs which at present is
the largest survey of homogeneous abundances in globular cluster giants. Already
21 GCs (soon to become 24 with the addiction of NGC~362, NGC~4833, and NGC~6093)
are analyzed and this offers a unique opportunity. In a forthcoming paper we
will apply the method developed here to derive homogeneous estimates of the
fraction of first and second generation stars in all the globular clusters of
this sample, to compare them with previously derived values from spectroscopy
and/or photometry, and to study their relationship with global cluster
properties.

\section{The comparison sample of field stars}

To assemble the comparison sample of field stars we searched the literature for
the most recent studies based on high resolution spectroscopy providing Na
abundances for a large number of stars of different metallicities. Our method 
to estimate the fraction of stars of different generations in GCs is based on
comparing the location of cluster and field stars in the Fe-Na plane; hence, the
main requisites are homogeneity of measurements and a good sampling in this
plane to highlight possible outliers. In particular, our aim was to secure
a good coverage in [Fe/H] over the range in metallicity spanned by the bulk of
GCs.

\subsection{Our approach: problems and methods}

About twenty abundance analysis with these characteristics can be easily picked
up in the past 15 years, with a number of stars analized ranging from 23
metal-poor giants in Johnson (2002) to more than 1000 FGK dwarfs surveyed with
HARPS in Adibekyan et al. (2012). 

Obviously, independent studies designed to study different stellar populations
to deal with different astrophysical problems involve a series of different
methods and assumptions that in turn may affect the requirement of homogeneity.
Among these we may include different temperature scales, based on photometry or
derived from the spectra, differences in the adopted reference solar abundances,
or in the scales of atomic parameters. The latter is less important for the 
Na~{\sc i} lines most used in abundance analysis, whose $gf$ are very well known
and homogeneously used, but may represent a source of discrepancy concerning Fe,
especially when coupled to different temperature scales.
Another potentially dangerous source of offset among different studies are the
corrections for departures from the LTE assumptions, in particular for Na. This
problem may become relevant especially when abundances derived for warm dwarfs
are compared to those obtained in metal-poor, cool stars of low gravity, where
the NLTE effects are stronger (see e.g. Baum\`uller et al. 1998, Gratton et
al. 1999, Korotin and Mishenina 1999, Mashonkina et al. 2000, Takeda et al.
2003, Gheren et al. 2004, Andrievsky et al. 2007, Lind et al. 2011, and
references therein). Moreover, different groups adopt different prescriptions
for NLTE corrections.

Previous extensive compilations of literature data, namely those by Venn et al. 
(2004; 821 stars), and by Soubiran and Girard (2005; 743 stars), were designed
to deal with specific problems (mainly related to the study of Galactic stellar
populations) and are not very  suitable to our purposes for various reasons. 
Venn et al. did not try to homogenise the abundance data apart from atomic data
for a few neutron-capture elements (Y, Ba, and Eu). Moreover, in their sample
they did not include the study by Gratton et al. (2003, hereinafter G), that we
will adopt as our reference sample (see below, Section 2.2).  Soubiran and
Girard used stars in common among their 11 selected samples to  operate an
homogeneisation of the total sample. However, their study was focused on the
interface between thick and thin disks, thus they restricted the metallicity
range to stars more metal rich than [Fe/H]$=-1.3$ dex, also to avoid the larger
scatter observed at low metallicities. Moreover, to further reduce the observed
dispersion, they limited their sample to stars hotter than 4500 K. However, one
of our aims is to provide a good sampling also in the low metal abundance
regime; since our main purpose is to build a sample to be compared mainly to
giant stars in rather metal-poor objects such as the GCs, the quoted existing
compilations are unsuitable (apart from missing recent studies performed after
2005).

\begin{figure*}
\centering
\includegraphics[bb=19 154 574 593, clip, scale=0.82]{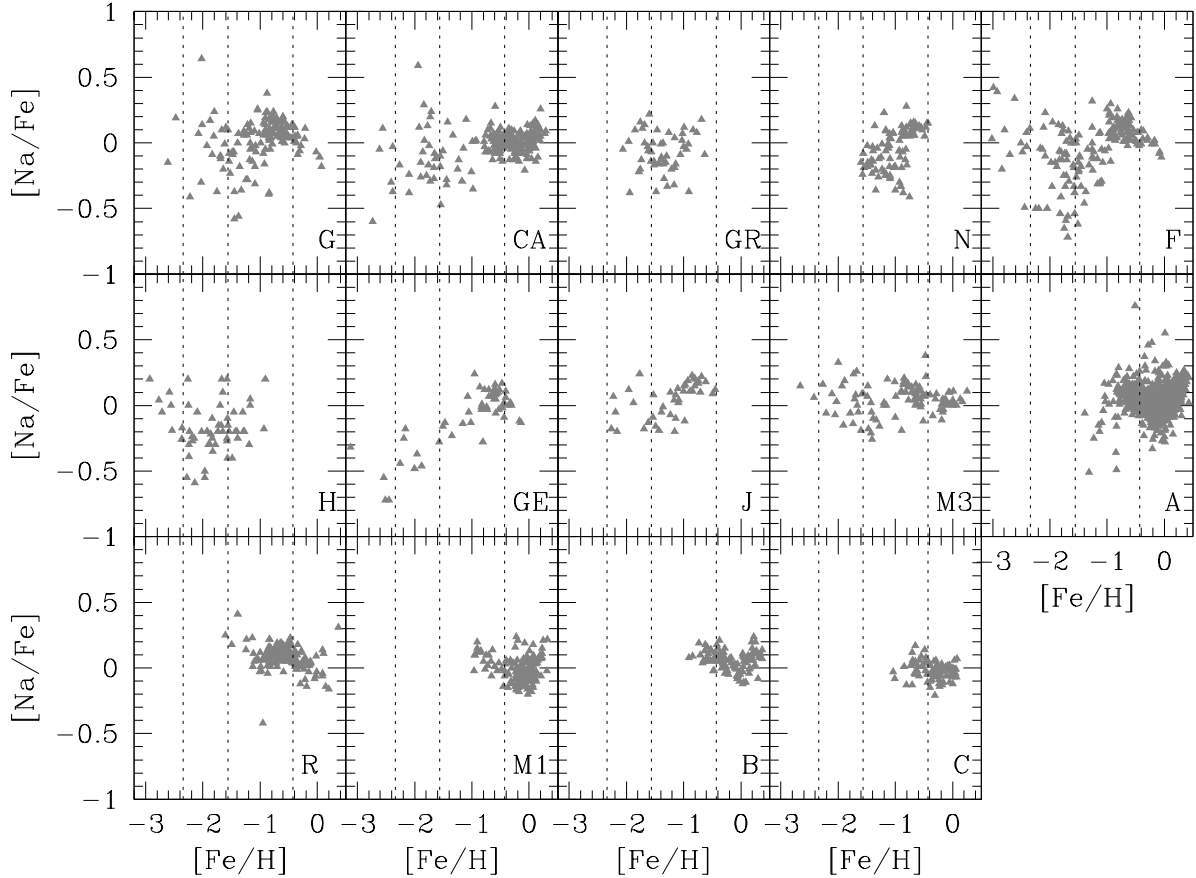}
\caption{[Na/Fe] abundance ratios as a function of [Fe/H] for the 14 samples
selected from the literature. The code-labels are listed in
Tab.~\ref{t:sample}. The [Na/Fe] and [Fe/H] values are those of the original
papers. The leftmost and rightmost dotted line indicate the metallicity range of
the bulk of GCs in the FLAMES survey (see e.g. Carretta et al. 2009a,b). The
line at [Fe/H]=-1.56 dex shows the metallicity of NGC~6752, our test case.}
\label{f:nafe14}
\end{figure*}

From our literature search, we eventually selected 14 studies, according to
criteria defined below in Section 2.4 . In Fig.~\ref{f:nafe14} we show the ratio
[Na/Fe] as a function of the metallicity [Fe/H] for all these samples, using the
original values in each paper. The labels refer to the sample coding listed in
Tab.~\ref{t:sample}. In each panel we indicate with dotted lines at
[Fe/H]$=-2.34$ dex and $-0.43$ dex the metallicity range spanned by the large
sample of GCs analysed by Carretta et al. (2009a,b) in the FLAMES survey of the
Na-O anticorrelation in globular clusters. The middle value, at [Fe/H]$-1.56$
dex, is the average metallicity (see Carretta et al. 2009c) of NGC~6752, that we
will use as our test case. 

\begin{table*}
\centering
\scriptsize
\caption{Mean properties of the samples}
\setlength{\tabcolsep}{1.5mm}
\begin{tabular}{llllllllllllll}
\hline
       & nr.   & with& min.   & max.   & evol.     & LTE or & pop.1     & pop.2       & pop.3   & comm.   & mean off.     & mean off.     &ref.                      \\
       & stars & Na  & [Fe/H] & [Fe/H] & stage     & NLTE   &           &             &         & stars   & [Fe/H]        & [Na/H]        &                          \\
       &       &     &        &        &           &        &           &             &         &         &	          &               &                          \\
\hline
       &       &     &        &        &           &        &           &             &         &         &	          &               &                          \\
G      & 150   & 147 & -2.61  & +0.07  & dwarfs,   & NLTE   & accret.   & dissip.     & thin    &         &               &               & Gratton et al. (2003)    \\
       &       &     &        &        & SGB       &	    & comp.     & comp.       & disk    &         &               &               &                          \\
       &       &     &        &        &           &        &           &             &         &         &	          &               &                          \\
CA     & 286   & 237 & -2.84  & +0.29  & dwarfs,   & NLTE   & halo      & thick       & thin    &  41     & $-$0.10 dex   &   +0.01 dex   & Carretta et al. (2000)   \\
       &       &     &        &        & giants    &	    &           & disk        & disk    &	  & $\sigma=0.07$ & $\sigma=0.08$ &                          \\
       &       &     &        &        &           &        &           &             &         &         &	          &               &                          \\
GR     &  58   &  49 & -2.31  & -0.63  & dwarfs,   & NLTE   & halo      & thick       &         &  14	  & $-$0.13 dex   & $-$0.02 dex   & Gratton et al. (2000)    \\
       &       &     &        &        & giants    &	    &           & disk        &         &	  & $\sigma=0.05$ & $\sigma=0.09$ &                          \\
       &       &     &        &        &           &        &           &             &         &         &	          &               &                          \\
\hline
\hline
       &       &     &        &        &           &        &           &             &         &         &	          &               &                          \\
       &       &     &        &        &           &        &           &             &         &         &	          &               &                          \\
N      &  94   &  94 & -1.60  & -0.63  & dwarfs    & LTE    & halo	& thick       &         &  35	  & $-$0.05 dex   & $-$0.06 dex   & Nissen \& Schuster (2010)\\
       &       &     &        &        &           &	    &		& disk        &         &	  & $\sigma=0.07$ & $\sigma=0.04$ &                          \\
       &       &     &        &        &           &        &           &             &         &         &	          &               &                          \\
F      & 178   & 174 & -3.01  & -0.05  & dwarfs,   & LTE    & halo      & thick       & thin    & 110	  &   +0.05 dex   &   +0.06 dex   & Fulbright (2000)         \\
       &       &     &        &        & giants    &	    &           & disk        & disk    &	  & $\sigma=0.06$ & $\sigma=0.05$ &                          \\
       &       &     &        &        &           &        &           &             &         &         &	          &               &                          \\
H      &  59   &  59 & -2.93  & -0.91  & giants    & LTE    & halo      &             &         &  32	  &   +0.05 dex   &   +0.27 dex   & Hanson et al. (1998)     \\
       &       &     &        &        &           &	    &           &             &         &	  & $\sigma=0.12$ & $\sigma=0.16$ &                          \\
       &       &     &        &        &           &        &           &             &         &         &	          &               &                          \\
GE     &  55   &  55 & -3.12  & -0.14  & dwarfs    & NLTE   & halo      & thick       & thin    &  23	  & $-$0.03 dex   & $-$0.00 dex   & Gehren et al. (2006)     \\
       &       &     &        &        &           &	    &           & disk        & disk    &	  & $\sigma=0.05$ & $\sigma=0.07$ &                          \\
       &       &     &        &        &           &        &           &             &         &         &	          &               &                          \\
J      &  43   &  42 & -2.99  & -0.45  & turnoff,  & LTE    & halo	& thick       &         &  18	  &   +0.00 dex   & $-$0.03 dex   & Jonsell et al. (2005)    \\
       &       &     &        &        & SGB       &	    &		& disk        &         &	  & $\sigma=0.08$ & $\sigma=0.06$ &                          \\
       &       &     &        &        &           &        &           &             &         &         &	          &               &                          \\
M3     & 100   & 100 & -2.66  & +0.25  & dwarfs,   & NLTE   & halo      & thick       & thin    &  34	  & $-$0.04 dex   & $-$0.11 dex   & Mishenina et al. (2003)  \\
       &       &     &        &        & giants    &	    &           & disk        & disk    &	  & $\sigma=0.09$ & $\sigma=0.17$ &                          \\
       &       &     &        &        &           &        &           &             &         &         &	          &               &                          \\
A      &1111   &1110 & -1.39  & +0.55  & dwarfs    & LTE    & halo      & thick       & thin    &  24	  & $-$0.07 dex   & $-$0.08 dex   & Adibekyan et al. (2012)  \\
       &       &     &        &        &           &	    &           & disk        & disk    &	  & $\sigma=0.07$ & $\sigma=0.11$ &                          \\
       &       &     &        &        &           &        &           &             &         &         &	          &               &                          \\
R      & 176   & 164 & -1.98  & +0.37  & dwarfs    & LTE    & halo      & thick       & thin    &  43	  & $-$0.07 dex   & $-$0.05 dex   & Reddy et al. (2006)      \\
       &       &     &        &        &           &	    &           & disk        & disk    &	  & $\sigma=0.08$ & $\sigma=0.08$ &                          \\
       &       &     &        &        &           &        &           &             &         &         &	          &               &                          \\
M1     & 142   & 142 & -0.96  & +0.32  & dwarfs    & LTE    & thick     & thin        & Her mov.&  10	  &   +0.01 dex   &   +0.04 dex   & Mishenina et al. (2011)  \\
       &       &     &        &        &           &	    & disk	& disk        & group   &	  & $\sigma=0.07$ & $\sigma=0.06$ &                          \\
       &       &     &        &        &           &        &           &             &         &         &	          &               &                          \\
B      & 102   & 102 & -0.91  & +0.37  & dwarfs    & LTE    & thick	& thin        &         &   8	  & $-$0.09 dex   & $-$0.05 dex   & Bensby et al. (2005)     \\
       &       &     &        &        &           &	    & disk	& disk        &         &         & $\sigma=0.14$ & $\sigma=0.07$ &                          \\
       &       &     &        &        &           &        &           &             &         &         &	          &               &                          \\
C      &  90   &  81 & -1.04  & +0.08  & dwarfs    & LTE    & thin	&             &         &   5	  &   +0.01 dex   &   +0.06 dex   & Chen et al. (2000)       \\
       &       &     &        &        &           &	    & disk	&             &         &         & $\sigma=0.04$ & $\sigma=0.05$ &                          \\
\hline
\end{tabular}
\label{t:sample}
\end{table*}

In Tab.~\ref{t:sample} we listed a one/two letter coding for each sample,
together with a summary of a few properties of the samples. The original studies
cover all the major Galactic stellar populations, thin and thick disks, and
halo. Most objects are dwarf stars, but also subgiants and  giants are
represented in some samples. In column 7 we indicated whether the abundance
analysis for Na was made under the LTE assumption or corrections for NLTE were
applied (references in the latter case can be found in the original papers).

Aided by the reference lines at different metallicities, from 
Fig.~\ref{f:nafe14} we first note that different samples occupy more or less the
same position, indicating that offsets among different analyses are not severe.
Second, there is some evidence of increased scatter in [Na/Fe] ratios at low
metallicity, in particular between [Fe/H]$\sim -2.3$ dex and  [Fe/H]$\simeq
-1.5$ dex. This occurrence, already noted in several studies, can be
appreciated  particularly well in a few samples (Gratton et al. 2003, Fulbright
2000, and Hanson et al. 1998).

In the following analysis, however, we chose to switch to the plane [Na/H] $vs$
[Fe/H] for two main reasons. The first was simply to decouple Na and Fe,
avoiding that errors in Fe reflect on both axis. The second is to fully exploit
the different sites of nucleosynthesis for Na. The main production of Na is
considered to be hydrostatic carbon burning in massive stars, a primary mechanism
(see e.g. Woosley and Weaver 1995). Since Na is an odd-Z element, its production
is a strong function of the neutron density, which is a function of the initial
stellar metallicity (Truran and Arnett 1971, Woosley and Weaver 1995). A simple
linear relation between [Na/H] and [Fe/H] is then expected. However,
proton-captures on $^{22}$Ne in H-burning at high temperature result in the
synthesis of Na able to explain the excess of Na observed in GC stars
(Denisenkov and Denisenkova 1989). Moreover, Langer et al. (1993) proposed that
synthesis of $^{23}$Na on the abundant isotope $^{20}$Ne may greatly contribute
to the Na production via proton-capture. Hence, we can exploit these different
production mechanisms for Na to our purposes.

Therefore, the [Na/H]-[Fe/H] plane could be considered one of the best
diagnostics to separate the stars with chemical signature from plain supernova
nucleosynthesis (both in the Galactic field populations and in globular
clusters) from stars where the proton-capture reactions have acted to modify the
original chemical imprinting. Any excess of [Na/H] in GC stars, with respect to
the linear behaviour in field stars at any given metallicity, can be interpreted
as a signature of second generation stars.

\begin{figure*}
\centering
\includegraphics[bb=19 154 574 593, clip, scale=0.82]{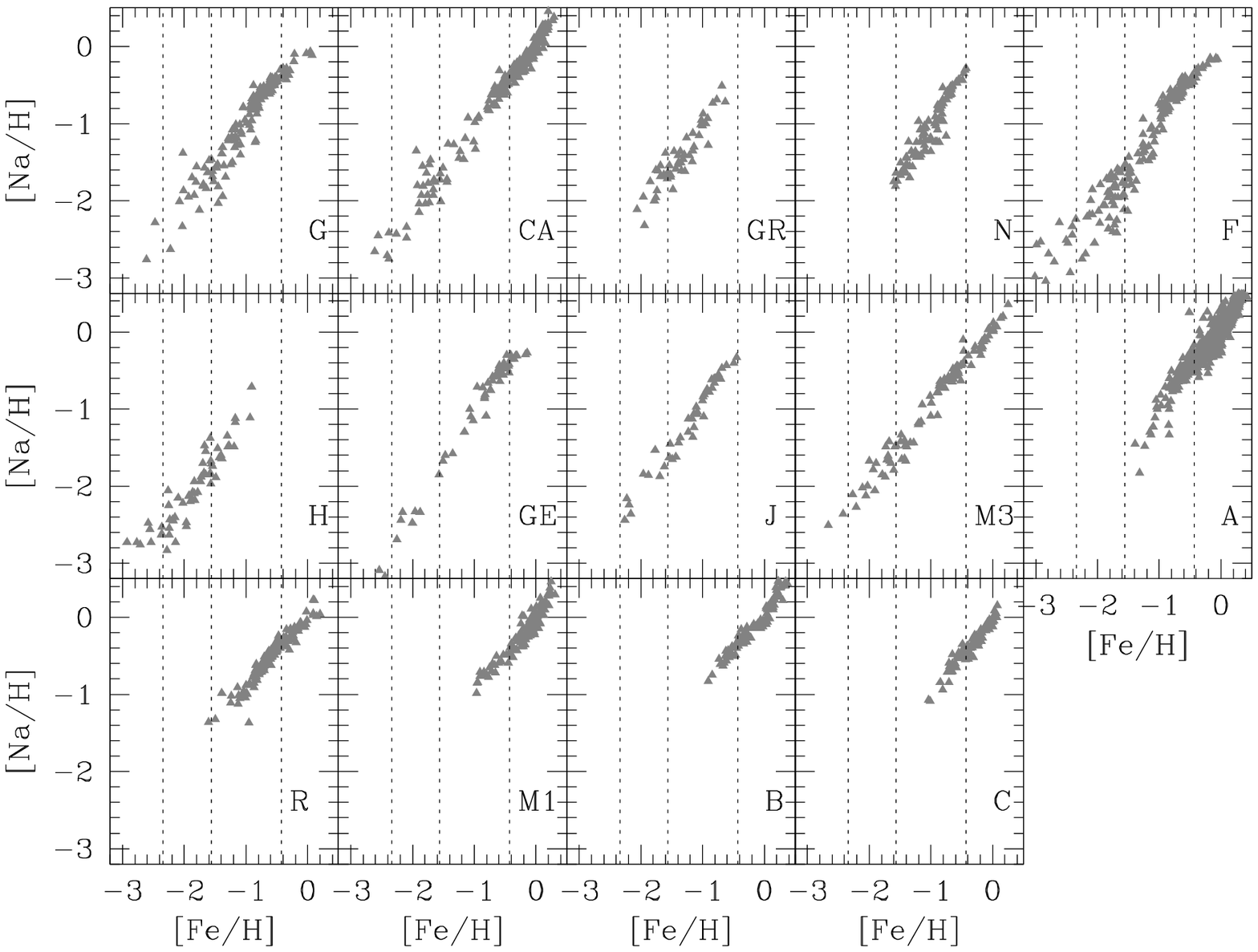}
\caption{As in Fig.~\ref{f:nafe14} but for the [Na/H] ratios as a function of 
[Fe/H] for the 14 samples selected from the literature.}
\label{f:nah14}
\end{figure*}

The run of the [Na/H] ratios as a function of metallicity is shown in
Fig.~\ref{f:nah14} for all the 14 datasets used to build our total comparison
sample of field stars. Possible systematics among them due to the different
scales and methods adopted in the original studies were treated by mean of
offsets computed using stars in common to bring all the [Na/H] and [Fe/H] 
values onto a homogeneous system defined by a reference sample, and 
corroborated by an extended reference sample, as discussed below.

\subsection{The reference sample: Gratton et al. (2003)}

Our privileged reference sample is the dataset of 150 field subdwarf and early
subgiant stars with accurate parallaxes analysed by Gratton et al. (2003;
hereinafter G, see Tab.~\ref{t:sample} for details). That work was based on high
resolution spectra for about 50 stars and on a collection of high quality
equivalent widths ($EW$s) from the studies by Nissen and Schuster (1997),
Prochaska et al. (2000), and Fulbright (2000).

There are several reasons to choose this set as reference sample:
\begin{itemize}
\item Adopted solar reference abundances for Na (6.21) and Fe (7.54) are the
same as used in the FLAMES survey of Na,O abundances in GCs (including
homogeneous abundances for almost 25 clusters);
\item in like vein, the atomic parameters scale and line list are the same for
field stars in G and cluster stars; as said above, this is not very relevant for
Na, but it could be a plus concerning the iron abundances;
\item the grid for corrections from NLTE effects used for Na (from Gratton et
al. 1999) is the same also adopted for Na abundances in GC stars of the above
extensive survey;
\item the G sample homogeneously covers all the metallicity range relevant for 
GCs, going from [Fe/H]$=-2.61$ dex up to $+0.07$ dex.
\end{itemize}

Overall, taking into account the obvious differences due to dealing with nearby
field stars rather than distant cluster stars, the G sample looks like the best
reference sample to define a common system for our total comparison sample.

All the main stellar populations in the Galaxy are sampled in G: thin disk,
thick disk, and halo. However, Gratton et al. (2003) did not adopt this
classical classification. Using the orbital characteristics of the
sample they distinguished stars belonging to a dissipative component (as
proposed e.g. by Eggen et al. 1962, and including objects from the classical
populations of halo and thick disk) and stars of an accretion component, as in
the scenario proposed by Searle and Zinn (1978) for halo GCs. This
classification scheme is becoming more and more used, thanks to large scale
survey such as the SDSS that stimulated many observational and theoretical works
(see e.g. Carollo et al.  2013 and Zolotov et al. 2009 for the dual halo
nature). The relevance of the accretion versus dissipative-collapse component
will be clear when discussing the outliers in the field star total
sample (see Section 3 below).

A possible shortcoming of this reference sample is that it does not include 
giant stars, that are the vast majority of stars usually observed in distant
GCs. We will see that this is not a source of concern because we can tie
together the worlds of giants and dwarfs by using the sample of unevolved stars
observed in globular clusters, in particular in our test case NGC~6752.

However, to ensure the reliability of the comparison sample of field stars, we
defined an {\it extended reference sample} that (i) provides a larger number of
stars for the cross-match with other studies for the homogeneisation of the
different datasets, and (ii) includes a number of giants to check that in the
[Na/H]-[Fe/H] plane they occupy the same region of dwarfs, despite e.g. the
different amounts of correction for NLTE on Na.

\subsection{The extended reference sample}

To build up our extended reference sample we considered the study of about 300
field stars by Carretta et al. (2000; hereinafter code CA), in the metallicity
range from [Fe/H]$\simeq -2.8$ dex to [Fe/H]$\simeq +0.3$ dex (see
Tab.~\ref{t:sample}) encompassing the metal abundance distribution of GCs. CA
analysed or reanalysed stars using $EW$ from high resolution spectra using
homogeneous methods. The main advantages of adding this study are that both
giants and dwarfs are included in the sample, the corrections for NLTE effects
on Na are from the same grid of Gratton et al. (1999) used by G, and more than
40 stars are in common with G, allowing to compute precise offsets to compensate
for the different scales of atmospheric parameters and linelists. Finally, the
CA sample includes the reanalysis of the extensive set of disk  stars by
Edvardsson et al. (1993); present in both the large compilations by Venn et al.
(2004) and Soubiran and Girard (2005), we preferred to use the sample as
reanalysed by CA to ease the homogeneisation of the final sample.

Adopting the CA sample we may check for possible systematic  differences between
dwarfs and giants due e.g. to the corrections for NLTE effects on Na. Owing to
large ionising fluxes and high densities, the corrections are usually less
important for unevolved stars than for giants.

\begin{figure}
\centering
\includegraphics[scale=0.42]{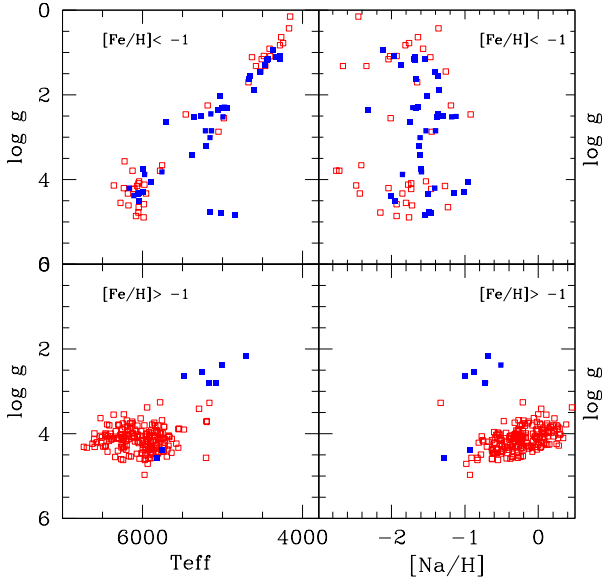}
\caption{Left panels: surface gravity as a function of effective temperature for
stars in the CA sample (open squares) and in the GR sample (blue filled
squares). In the upper panel are plotted only metal-poor stars ([Fe/H$<-1$ dex),
while in the lower panel stars have [Fe/H]$>-1$ dex. Right panels: the
surface gravity is plotted as a function of [Na/H], with the same symbols and
metallicity ranges used in the left panels. Na abundances are corrected for
departures from LTE according to Gratton et al. (1999) in both samples.}
\label{f:CAlogg}
\end{figure}

In Fig.~\ref{f:CAlogg} (left panels) we show the temperature-gravity diagram for
stars in the CA sample (open red squares), divided into two subsamples of
different metallicity at [Fe/H]$=-1$ dex. While the higher metallicity subsample
contains only dwarfs (being mostly based on the reanalysis of the Evdardsson et
al. dataset), at low metallicity both giants and unevolved stars are present.
The values of [Na/H], including corrections for NLTE effects from Gratton et al.
(1999), are plotted in the right panels as a function of the gravity. No offset
in Na abundances can be seen between low and high gravity stars.

To further reinforce our reference sample, we also added another study from the
group of Gratton and collaborators, including in our extended reference sample
the about 60 stars analysed by Gratton et al. (2000; hereinafter code GR, see
Tab.~\ref{t:sample} for further details). This is a set of stars mostly in the
metallicity range $-2\leq$[Fe/H]$\leq -1$ dex, with well determined evolutionary
phases from the main sequence up to the upper red giant branch.
Again, as in CA, temperature scale and linelist are different from those adopted
by G, but the NLTE corrections applied to Na are from the same grid used by G
and Ca.

In Fig.~\ref{f:CAlogg} stars of the GR sample are plotted as filled blue
squares. At low metallicity the sequence in the T$_{\rm eff}-\log g$ plane is
well defined, and a good sampling of all the evolutionary phases is evident.
No difference in the locus occupied by stars in the different phases is visible
in the right panels, even among the much more limited metal-rich subset.

Thus, we adopted the G sample as our favourite reference sample, that, in
combination with the CA and GR datasets, constitutes our extended reference
sample. These three sets are separated in Tab.~\ref{t:sample} from the
other samples.

\subsection{The final comparison sample}

From our literature search we retained only the studies with at least five
analysed stars in common with our reference sample G or more than 10 objects in
common with the total extended sample. Apart from the three already quoted
samples, the studies we used were: Hanson et al.
(1998, hereinafter code H), Fulbright (2000; F), Chen et al. (2000; C),
Mishenina et al. (2003; M3), Jonsell et al. (2005; J), Bensby et al. (2005; B), 
Gehren et al. (2006; GE), Reddy et al. (2006; R), Nissen and Schuster (2010; N),
Mishenina et al. (2011; M1), Adibekyan et al. (2012; A).
The code labels, the total number of stars and those with Na abundances, the
metallicity range, the main evolutionary stage, whether the Na abundance
analysis was or was not in LTE, and the main Galactic populations sampled in
each study are indicated in the first 10 columns of Tab.~\ref{t:sample}. In
column 11 the number of stars in common between each original study and the
reference sample G is listed.

By using these stars in common we computed the offsets in [Na/H] and [Fe/H],
applying, when required, a 2.5 $\sigma$-clipping with respect to the mean
offset. The resulting average values (in the sense: G minus other sample) are
listed in columns 12 and 13 together with the r.m.s. scatter about the mean. The
amount of these offsets is generally not very large, and scarcely significant,
given the associated scatters. 
An exception is the offset in Na measured for the H sample, 0.27 dex (from 25
stars after k$\sigma$-clipping). Na abundances in H are actually those obtained
by Pilachowski et al. (1996); by using their sensitivity Table 7, if the
discrepancy was due to uncorrect values of temperature and gravity, the required
differences would be of 386 K in T$_{\rm eff}$ or 2.7 dex in $\log g$, while the
observed mean offset are only -24 K (rms=105 K) and -0.06 dex (rms=0.42 dex),
respectively. Hence, we cannot offer at present an explanation; we only note
that a wide scatter in Na abundances was already claimed by Pilachowski et al.
(1996). Moreover, H is the only sample where no stars in common with G do
exists, but only 21 stars in common with CA and 11 with GR are available.
However, the application of the derived offsets brings the H measurements 
in better agreement with the other samples; the inclusion of the H sample is
valuable for reasons discussed below.

In absolute value, the mean offsets in temperature and gravity for the 13
samples are $|T_{\rm eff}|=49,\ \sigma=30$ K and $|\log g|=0.09,\ \sigma=0.06$
dex. Using the sensitivities of Na and Fe to variations in the
atmospheric parameters listed in the original papers, we verified that the
measured offsets with respect to the G scale translate into uncertainties not
exceeding  $\pm 0.05$ dex in [Na/H] (and similar amounts in [Fe/H]).

The final sample was obtained by eliminating all objects without Na abundance
and applying the proper offsets in [Na/H] and [Fe/H] to all stars in each
dataset.
Stars with multiple determinations were treated according to the following
criteria: (i) all stars in the G dataset were retained; (ii) for multiple
measurements with no star from G, those in CA or GR samples were retained, if
present, averaging their values if a star was measured in both sample; (iii)
stars measured in more than one of the other samples were averaged.

The total final sample consists of 1891 individual, unique stars on the scale
defined by the G sample. Among these objects, 147 are from G, 200 from CA, 32
from GR, 39 from N, 30 from F, 26 from H, 14 from GE, 9 from J, 21 from M3, 1014
from A, 88 from R, 80 from M1, 40 from B, 49 from C, and 102 are averaged values
from different studies. The assembled dataset covers the metallicity range from
[Fe/H]$=-3.15$ dex up to [Fe/H]$=+0.48$ dex. Restricting the range to the
interval $-2.36\leq$[Fe/H]$\leq -0.33$ dex where are located the GCs studied in
the FLAMES survey by Carretta et al. (2009a,b), and the bulk of GCs in general,
the field star sample includes 804 stars. A good fraction of stars, especially
at high metallicity, is from the sample A, devoted to the chemical analysis of
dwarf stars from the HARPS GTO planet search program. Objects outside the
metallicity range most relevant for GCs could well be rejected, but we retained
all the same in our final sample because it could be used in future to study the
Galactic chemical evolution of Na using a large and homogeneous dataset.

\begin{table*}
\centering
\scriptsize
\caption{Final sample}
\begin{tabular}{llcccclrll}
\hline
n      & HD/BD/G    &$\log g$& T$_{\rm eff}$ & [Fe/H] & [Na/H] & sample & HIP & R.A.(2000) & DEC.(2000) \\
\hline
 0001  &   HD 224930&  4.32  &5357 & $-$0.90& $-$0.67 & G    & 171 & 00 02 10.15644 &+27 04 56.1304 \\
 0002  &   HD 3567  &  4.16  &6087 & $-$1.22& $-$1.50 & G    &3026 & 00 38 31.94747 &-08 18 33.3952 \\
 0003  &   HD 3628  &  4.01  &5704 & $-$0.21& $-$0.10 & G    &3086 & 00 39 13.26416 &+03 08 02.1311 \\
 0004  &  CD-35 0360&  4.53  &5048 & $-$1.15& $-$1.19 & G    &5004 & 01 04 06.12974 &-34 40 28.9509 \\
 0005  &   HD   6582&  4.46  &5322 & $-$0.87& $-$0.79 & G    &5336 & 01 08 16.39470 &+54 55 13.2264 \\
\hline
\end{tabular}
\label{t:uniche}
\begin{list}{}{}
\item[1-] the code AV in the sample column means that the star was observed in 
more than one sample and the listed values of [Fe/H] and [Na/H] are averages.
\end{list}

\end{table*}

In Tab.~\ref{t:uniche} we list an increasing order number, the star name,
gravity and temperature, the [Fe/H] and [Na/H] values corrected to the G scale,
the original sample code (AV means that the Fe and Na values from different
samples were averaged), the Hipparcos name (when available; else 000000 is
listed), the right ascension and declination, referred to the 2000 equinox, as
listed in SINBAD. This table will be available only in electronic form at the
CDS database. Here we only present a few lines as an indication of its content.

\begin{figure}
\centering
\includegraphics[scale=0.40]{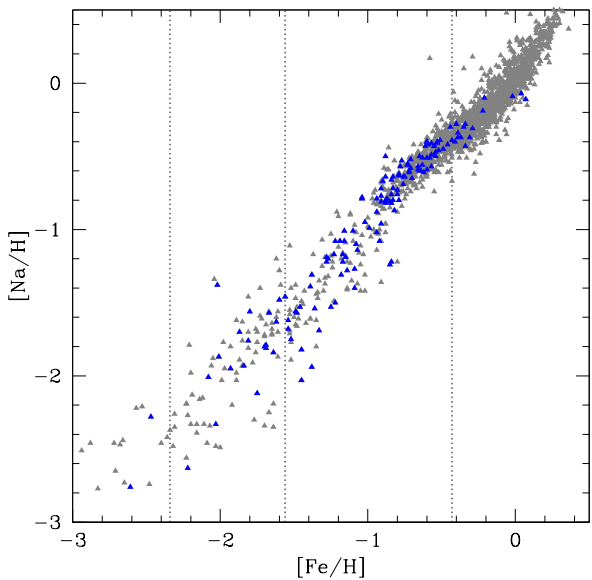}
\caption{[Na/H] ratios as a function of the metallicity in our final sample,
after correcting for the offsets in Na and Fe with respect to G. Stars in the G
sample are evidenced as blue symbols.}
\label{f:uniche}
\end{figure}

The [Na/H] values of this final sample are plotted as a function of metallicity
in Fig.~\ref{f:uniche}. Stars of the G sample are evidenced in this figure to
show how the reference sample well covers all the range in [Fe/H] of interest
for GCs. 

\section{Outliers in the field comparison sample}

There are stars in Fig.~\ref{f:uniche} that clearly deviate from the almost
linear relation between Na and Fe. This occurrence is already seen in
Fig.~\ref{f:nafe14} (as an increased scatter at low metallicity) and in
Fig.~\ref{f:nah14} in the original samples, in particular in G, F, and H. As a
consequence, this behaviour cannot be a spurious effect due, e.g., to the
corrections applied to the original [Na/H] and [Fe/H] values to bring them onto
the G system.

Since our method to derive the fraction of second generation stars is based on
the $excess$ of Na with respect to a baseline provided by field stars, it is
important to better understand the nature of these outliers, especially because
the deviation from linearity seems to become more and more relevant just at the
metal abundance of our test case NGC~6752 ([Fe/H]$\simeq -1.5$). 

\begin{figure}
\centering
\includegraphics[bb=105 150 438 709, clip, scale=0.62]{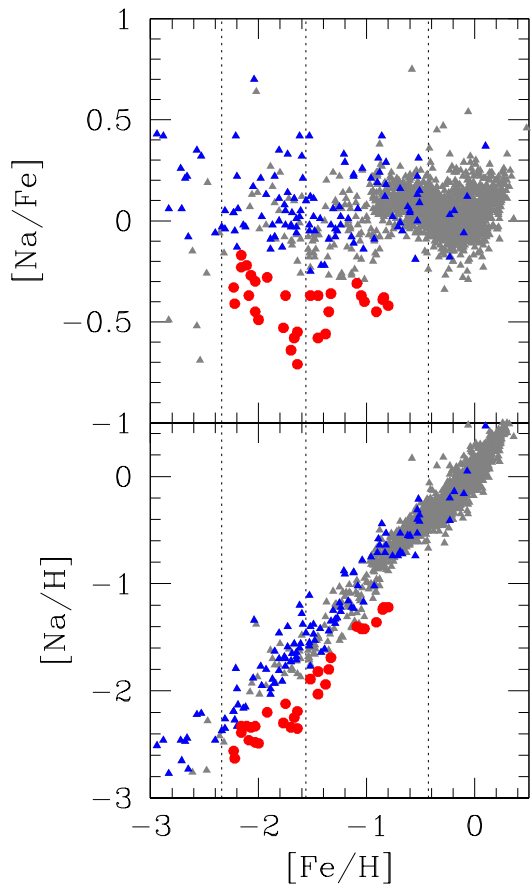}
\caption{Upper panel: [Na/Fe] ratios as a function of metallicity for stars in
the final homogeneized sample (grey triangles for dwarfs and blue triangles
for giants). Larger filled circles represent stars individuated as outliers
with respect to the bulk of the distribution (with no distinction between
dwarfs and the few giants). Dotted line have the same meaning as in previous
figures. Bottom panel: the same, but for the ratios [Na/H].}
\label{f:outliers}
\end{figure}

We selected in Fig.~\ref{f:outliers} those stars that seems to deviate in Na  at
a given metallicity from the bulk of the distribution, to understand if there is
some common feature that may explain the lower than average Na abundances.

The selection of these outliers was made by eye, but is far from subjective and
was checked by a linear fit in the [Na/H]-[Fe/H] plane of all the 1891 stars in
the final sample. The dispersion in [Na/H] around this fit is 0.123 dex, and we
checked that all candidate outliers have a sodium abundance from
$\sim 2$ up to 5.8$\sigma$ lower than that derived from the fit at any given
metallicity\footnote{The low dispersion around the best fit is driven by the
large number of stars in the high metallicity regime. Using only stars more
metal poor than [Fe/H]$=-0.3$ dex the r.m.s.scatter of the fit increases to
0.146 dex; in turn, the range of standard deviations spanned by the outliers
with low Na decreases, but is however in the interval $1.3 \div 4.8\sigma$}.

Among the 30 stars selected in Fig.~\ref{f:outliers}, a good fraction (10 stars,
30\%) is from the G sample. The majority belongs to the accretion component
defined by G (8 stars out of 10), 6 of them being found on retrograde orbits, 
as does the single star from GR included among the 30 selected stars. Six stars
from N (including 3 objects in common with G and another star in common with A)
are all of the low-$\alpha$ group defined in N (see also Schuster et al. 2012).

Another third of stars in the selected set is from F (8 stars were also in
common with H, and 2 are average values with H and M3). Most of them (7 out of
10) are on retrograde orbits.  Three other stars are unique objects from H, and
two of them were also found on retrograde orbits.  All stars in this group from
F are among the highest velocity stars in Fulbright (2002), who found for them a
different patterm of abundances with respect to  lower velocity stars. In
particular, the mean values for the [Na/Fe] and [Mg/Fe] were about
0.2 dex lower for the high velocity group (see Figure 6 in Fulbright 2002). 

One of the three low-$\alpha$ stars studied by Ivans et al. (2003; BD+80~245 =
HIP 40068) is included in this low-Na subset, as is the common proper motion
pair HD~134439/40 whose formation was associated by King (1997) to the
environment typical of dwarf spheroidals on the basis of the observed
underabundances in $\alpha$-elements. All three stars have large apogalactic
distances (about 20 kpc for HIP 40068, Fulbright 2002; more than 40 for the pair
HD~134439/40, King 1997).

It is clear that the sequence of stars selected with low Na abundances in
Fig.~\ref{f:outliers} is predominantly composed by objects whose origin is
attributed to accretion in the dichotomy of the Galactic halo amply studied by
several groups (e.g. Hanson et al. 1998, Fulbright 2000, 2002, Stephens and
Boesgaard 2002, Gratton et al. 2003, Nissen and Schuster 2010, Schuster et al.
2012 and references therein). It is convenient to adopt for these stars the
terminology used by Nissen and Schuster (1997): these objects are better seen as
iron-enriched than as $\alpha-$ or Na-poor. They formed in regions where the
usual nucleosynthesis by type II SNe produced the known amount of
$\alpha-$elements and Na; however, the original fragments were probably small,
not much dense and with a slow star formation rate. Hence, the low-$\alpha$ (Na)
component received chemical enrichment from both SNe II and SNeIa owing to the
slower chemical evolution timescales. The increased iron yields acted to dilute
the overabundance of some products from the SNe II down to the observed level.

By contrast, the bulk of stars with higher Na (the high-$\alpha$ group in Nissen and
Schuster 1997,2010,2011, Schuster et al. 2012; the dissipative component in G)
was generated from gas which experienced much more rapid chemical evolution,
preserving the abundances of Na and $\alpha-$elements. This component may
have been  formed in more massive primordial fragments or participated to the
early collapse of the proto-Galactic cloud. In the [Na/H]-[Fe/H] plane, the
latter component represents the bulk of our sample, 89\% of stars in the
metallicity range $-2.3 <$[Fe/H]$<-0.8$ dex is Na-normal.

\section{Application of the method: the case of NGC~6752}

After setting the stage, it is time to introduce our main players, the multiple
populations of globular clusters. We selected NGC~6752 to test the method.
This is one of the closest GC, often its stars are observed as calibrators in
many spectroscopic studies, and it has been the target of several accurate
abundance analyses and photometric studies as well. Among many others,
nitrogen abundances were studied by Yong et al. (2008), whereas a detailed study
of the Mg isotopes was performed by Yong et al. (2003). Recently, three discrete
stellar populations on the RGB were recognized first by Carretta et al. (2011)
by  using Grundahl's Str\"omgren photometry from Calamida et al. (2007);
afterward, their different chemical composition was studied by Carretta et al.
(2012). The discreteness of the photometric sequences was confirmed by
Milone et al. (2013) with a larger set of photometric observations.

In the present context, we are interested in abundances of proton-capture
elements, in particular Na. One of the largest sample was analyzed in Carretta
et al. (2007), where Fe, Na, O abundances were derived for about 150 giants
using the multifiber FLAMES spectrograph. These data were complemented by Al,
Mg, and Si abundances in Carretta et al. (2012). Smaller samples, but generally
based on higher resolution and high S/N spectra, are from Yong et al. (2005,
2008) and Grundahl et al. (2002). We will mainly employ the samples from
Carretta et al. to better exploit the homogeneity of data.

\begin{figure}
\centering
\includegraphics[scale=0.40]{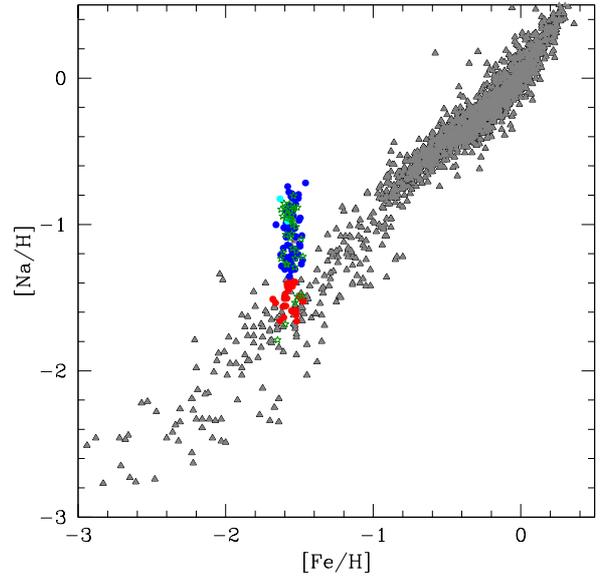}
\caption{Stars in NGC~6752 from Carretta et al. (2007) compared to our final 
sample (grey triangles). In NGC~6752 different symbols represent stars with
measured O, Na abundances of the primordial P component (red circles), the
intermediate I component (blue circles), and the extreme E component
(cyan circles) defined as in Carretta et al. (2009a). Green open star
symbols indicate cluster stars with only Na abundance measured.}
\label{f:nah6752}
\end{figure}

The stars in NGC~6752 from Carretta et al. (2007) are compared to field stars in
the present final sample in Fig.~\ref{f:nah6752} to illustrate our method.
Different symbols indicate the P, I, and E components as defined in Carretta et
al. (2009a) by using stars with both Na and O abundances. Cluster stars with
only Na measured are also indicated (open star symbols).

Even from this large scale figure it is evident that about 2/3 of stars in
NGC~6752 stand clearly off the distribution of field stars, implying that these
are actually stars with chemical composition modified by processes involving
proton-capture reactions and only restricted to the cluster environment.

The sample of cluster and field stars ultimately rest on the same scale 
concerning the line lists and NLTE corrections for Na. However, the effective
temperature scales adopted in Carretta et al. (2007) and in the reference sample
G are different, and  the presence of the low-Na field stars mainly from the
accretion component may generate a doubt on the vertical normalization. To check
this issue and nail down the samples we used stars with pure nucleosynthesis
from SNe, both in field and in NGC~6752, as follows.

\begin{figure}
\centering
\includegraphics[scale=0.40]{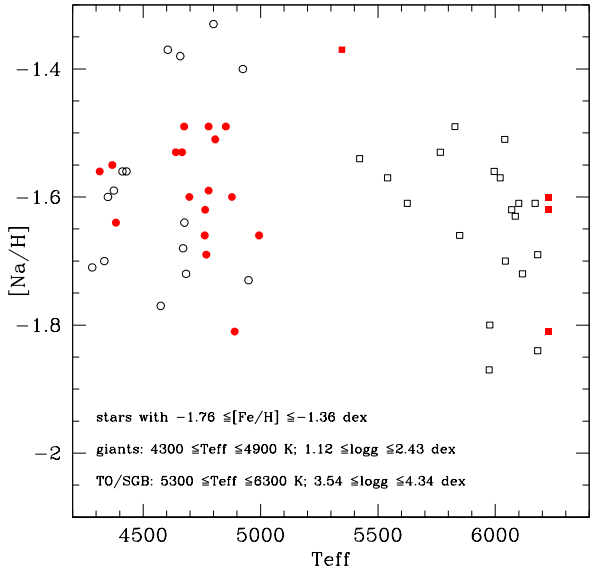}
\caption{[Na/H] ratios as a function of effective temperature to test the
concordance between field and cluster stars with nucleosynthesis from SNe.
Filled symbols are stars in NGC~6752 from Carretta et al. (2005, 2007).
Empty symbols are field stars in the present final sample. Circles indicate
giants and squares represent dwarf/subgiant stars.}
\label{f:inchioda}
\end{figure}

We selected a range of $\pm 0.2$ dex in [Fe/H] centered on the average
metallicity of NGC~6752 ([Fe/H]$=-1.56$ dex, Carretta et al. 2009c). We then
plotted in Fig.~\ref{f:inchioda} as a function of the effective temperature 
the [Na/H] ratios of all stars from Carretta et al. (2007) with [Na/Fe]$< +0.05$
dex and the four most Na-poor scarcely evolved stars from Carretta et al.
(2005), three turn off stars and one subgiant\footnote{These are stars studied
in Gratton et al. (2001) and simply reanalyzed in Carretta et al. (2005) by 
using updated line list and damping parameters, used afterward for GC giants.}.
This set of cluster stars is then chosen purposedly to represent the primordial
stellar population in GCs, the one reflecting only nucleosynthesis from SNe.

From the present final sample we then selected stars in the same metallicity
range, and matching the temperature and gravity ranges of cluster stars.
These field stars are plotted in Fig.~\ref{f:inchioda} and blend in with cluster
stars very nicely. There is no systematic difference in the [Na/H] content.
Furthermore, the sensitivity of [Na/H] to a change of 50 K in T$_{\rm eff}$ is
only 0.04 dex (Carretta et al. 2007), and to bring the average [Na/H]$\sim-1.6$
dex down to about -2.3 (to match the sequence of low-Na stars at the metallicity
of NGC~6752) would require a change/error of 875 K in temperature, frankly
unrealistic.

\begin{figure}
\centering
\includegraphics[scale=0.40]{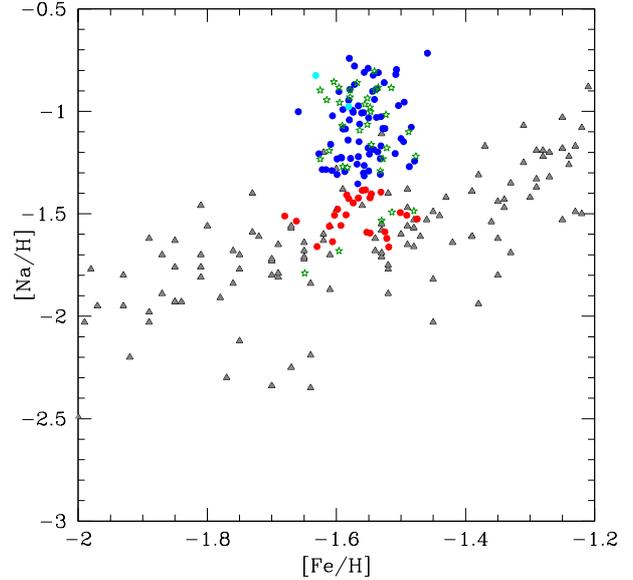}
\caption{Enlargement of Fig.~\ref{f:nah6752} centered on the average metallicity
of the globular cluster NGC~6752.}
\label{f:nah6752zoom}
\end{figure}

We could then safely proceed to the comparison between the distributions in 
[Na/H] of field and cluster stars. An enlargement of Fig.~\ref{f:nah6752} is
shown in Fig.~\ref{f:nah6752zoom}. From this Figure it is evident that the
division in components as defined in Carretta et al. (2009a) using both Na and O
abundances is approximatively still valid.

\begin{figure}
\centering
\includegraphics[scale=0.40]{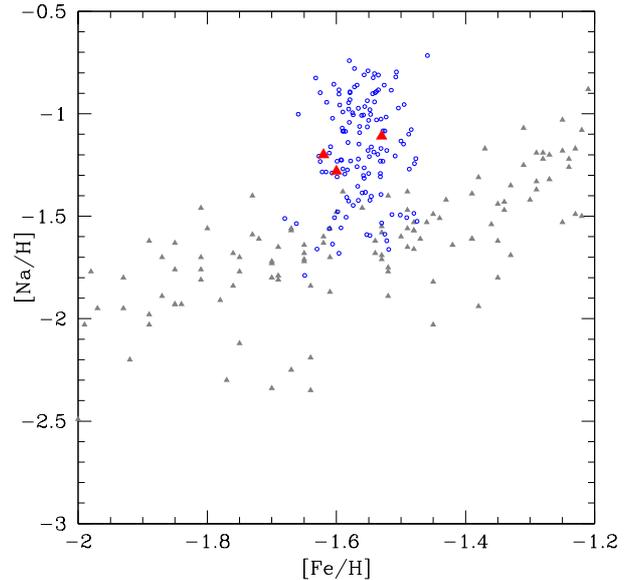}
\caption{[Na/H]-[Fe/H] plane in the metallicity range centered on NGC~6752 (open
blue symbols). Grey triangles are field stars and larger, red triangles
highlight three field stars with high Na abundance superimposed to stars in
NGC~6752.}
\label{f:interloper}
\end{figure}

A few outliers field stars can be seen superimposed to the stars in NGC~6752:
HD~93529, BD+54~1323, and BD+52~1601 (Fig.~\ref{f:interloper}). It is
however dubious that thay could represent new cases of second generation  GC
stars lost to the field. The first is from the H sample, is found on a
retrograde orbit and could be originated into a fragment accreted in the Galaxy,
maybe experiencing a peculiar chemical evolution. Travaglio et al. (2004)
reported BD+54~1323 among the stars suspected of AGB contamination, likely by
mass transfer from a past companion, since for this star Burris et al. (2000)
measured high values of light $s-$process elements. Finally, BD+52~1601 is
listed in the Fourth Catalog of Interferometric Measurements of Binary Stars
with a period of 9 hours.

\begin{figure}
\centering
\includegraphics[bb=25 149 361 717, clip, scale=0.52]{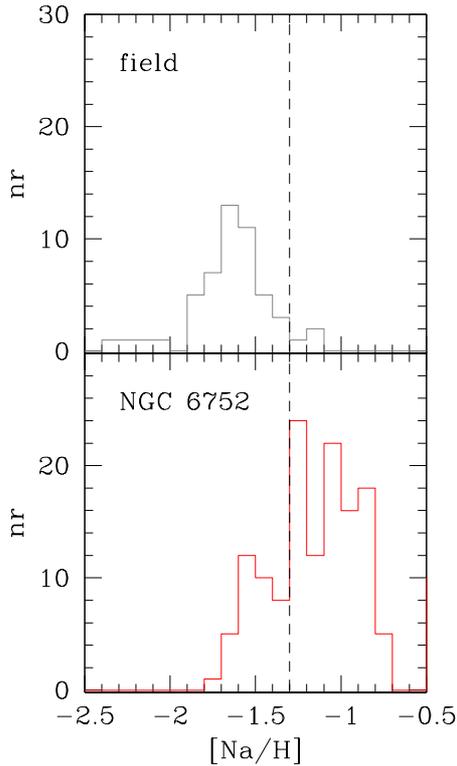}
\caption{Distribution of [Na/H] ratios in the metallicity range 
$-1.7<$[Fe/H]$<-1.4$ dex for field (upper panel) and cluster stars (lower
panel). The dashed line fixes the upper envelope of the field stars distribution
and the boundary between first and second generation stars in NGC~6752.}
\label{f:d2N6752}
\end{figure}

Excluding these three outliers, we estimated that the upper envelope of the
field star distribution can be put at [Na/H]$\sim -1.3$ dex
(Fig.~\ref{f:d2N6752}, dashed line), by comparing the distribution of field and
cluster stars in the metallicity range $-1.7<$[Fe/H]$<-1.4$ dex.
This value corresponds well to the 3-$\sigma$ upper deviation 
([Na/H]=-1.33 dex) from the field star relation obtained by eliminating
iteratively all the outliers around the fit [Na/H] vs [Fe/H] in the metallicity
range $-2.4 <$[Fe/H]$< -0.3$ dex, as suggested by the referee.

Adopting the limit in [Na/H] at -1.3 dex as the boundary between first and
second generation stars in NGC~6752, we counted 36 and 97 stars in the two
groups, respectively. On a total of 133 stars with Na abundances, we then
derived fractions of  $27\pm5\%$ and $73\pm7\%$ for the first and second
generation components in NGC~6752, where the associated errors are from the
Poisson statistics. These values are in perfect agreement with the fractions 
$27\pm5\%$ (P) and $73\pm9\%$ (I+E) obtained in Carretta et al. (2009a) by using
only the 98 stars with both Na and O abundances.

This new method is conceptually similar to that employed in Carretta et al.
(2009a), but the present variation allow us to make use of all stars with a
measured Na abundances, decreasing the statistical errors because it is usually
easier to measure abundances of elements that are enhanced (and not depleted) in
the proton-capture reactions and have much stronger line strengths. Obviously,
the present method is blind to the separation between the intermediate and
extreme components of second generation stars. 

\begin{figure}
\centering
\includegraphics[bb=18 145 367 712, clip, scale=0.52]{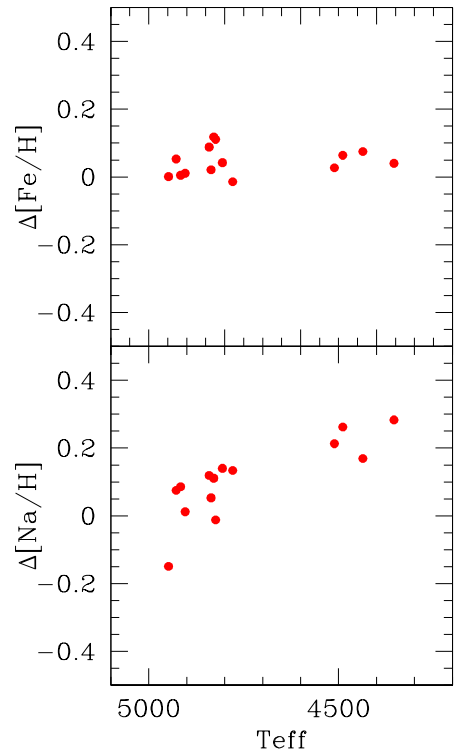}
\caption{Comparison of the differences in [Na/H] (lower panel) and [Fe/H] (upper
panel), in the sense Carretta et al. (2007) minus Yong et al. (2007), as a
function of the effective temperatures by Yong et al. (2005) for 14 stars
analyzed in both studies.}
\label{f:cfryong}
\end{figure}

For a further consistency check we used also the smaller sample of 37 stars
along the RGB in NGC~6752 presented in Yong et al. (2005). Na and Fe abundances
were derived from spectra generally of higher-resolution and S/N than most stars
in Carretta et al. (2007); these abundances were obtained using different scales
of atmospheric parameters and line lists. However, there are 14 stars in common
between the two samples: the differences in [Fe/H] and [Na/H] (in the sense
Carretta minus Yong) are shown in Fig.~\ref{f:cfryong}.

On average, the offset in [Fe/H] is not significant (0.046 dex, with an r.m.s.
scatter of 0.041 dex, 14 stars). However, a clear trend as a function of
temperature does exist, probably due almost entirely to the NLTE corrections for
Na (adopted from Gratton et al. 1999 in Carretta et al., neglected in Yong et
al.). They are relevant as both samples cover a large interval in luminosity and
temperature along the RGB in this nearby cluster. We then fitted a linear
regression through the data in the lower panel of Fig.~\ref{f:cfryong} and
derived the correction as a function of the effective temperatures by Yong et
al. required to shift their abundances on the system used in the present work.

\begin{figure}
\centering
\includegraphics[scale=0.40]{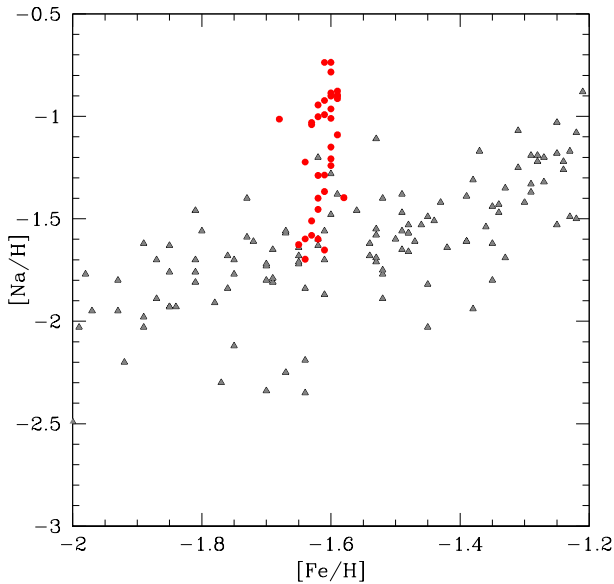}
\caption{As in Fig.~\ref{f:nah6752zoom}, but using for NGC~6752 the 37 stars
from Yong et al. (2005; red circles), corrected to the system of Na and Fe
abundances of the present work.}
\label{f:yong05}
\end{figure}

The separation between first and second generation at [Na/H]=-1.3 dex also in
the case of the corrected data by Yong et al. (Fig.~\ref{f:yong05}) returns
values in good agreement with our previous findings: we found that 11 stars out
of 37 ($30\pm 9\%$) have a primordial composition consistent with that of field
stars, whereas 26 stars ($70\pm 14\%$ of the sample) belong to the second
generation in NGC~6752.

\section{Final thoughs on the formation of the Galactic halo}

In the present work we discussed how second generation stars in globular
clusters may be selected from their excess in Na above the level established in
field stars. However, there are also a few field stars with chemical signatures
that indicate a very likely origin in GCs. Some begin to be  serendipitously
individuated thanks to their Na excesses (Carretta et al. 2010;  Ram\'irez et
al. 2012); in other cases focused studies purposedly search for second
generation GC stars lost to the field by looking at excess in the CN band
strength (Martell et al. 2011).

\begin{figure}
\centering
\includegraphics[scale=0.40]{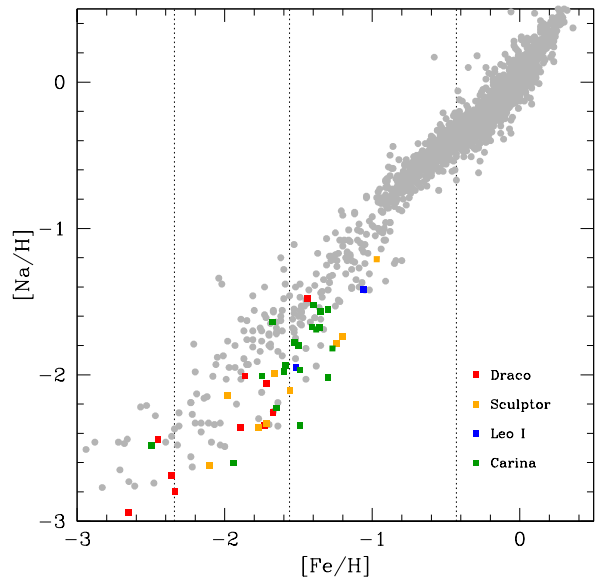}
\caption{Comparison of our final sample of field stars with stars in four dwarf
spheroidals: Draco (Cohen and Huang 2009; Shetrone et al. 2001), Sculptor (Kirby
and Cohen 2012, Shetrone et al. 2003, Geisler et al. 2005), Carina (Shetrone et
al. 2003, Venn et al. 2012, Koch et al. 2008), and Leo I (Shetrone et al.
2003).}
\label{f:dsph}
\end{figure}

Stars with typical chemistry of second generation stars in GCs are thus rather
easy to be recognized in the Galactic field. However, we have indirect evidence
that the bulk of the contribution of GCs to the halo must be in form of stars
bearing the primordial composition of first generation stars. This conclusion
stems (i) from the observed proportion of the stellar generations in GCs ($\sim
30\%$ primordial, $\sim 70\%$ polluted), (ii) from the evidence that the second
generation stars are formed from gas polluted by a fraction only of first
generation stars (the most massive), and (iii) from consequent theoretical
considerations and modeling (see e.g. Bekki et al. 2007, D'Ercole et al. 2008,
Vesperini et al. 2010, Schaerer and Charbonnel 2011 and references therein). All
these studies invoke a massive loss of almost all the primordial stellar
generation in early GCs (that were likely several times more massive than 
present-day GCs), to correctly reproduce the currently observed ratio of
multiple populations without the need of {\it ad hoc} initial mass functions
(IMFs). These stars would be not distinguished from normal halo field stars from
their chemical composition alone (this is just the  foundation of the present
approach), but, if present, they would be intermingled with the bulk of the
field stellar distribution.

However, we also signaled the presence of outliers, field stars with low Na
abundances, that are claimed by several studies to come from small fragments
accreted in the halo.

A not exaustive comparison of this component with the chemical pattern of a few
dwarf spheroidals (dSphs) is given in Fig.~\ref{f:dsph}. Most of the giants 
studied in dSphs have a tendency to lie below the main distribution of Galactic
field stars in the [Na/H]-[Fe/H] plane (see also Fig. 12 in Tolstoy et al. 2009,
with the more classical [Na/Fe] ratios). These stars well agree with the
accretion component of the Galaxy. Unfortunately, we have no stars in common
with the analyses in dSphs to bring them on our common system. However, the
same  line of reasoning as above suggest that different studies should be anyway
consistent  within a few hundreths of dex, else the temperatures would be wrong
by several hundreds of kelvin, an umpalatable option.

Hence, the results of this triple comparison (Galactic field stars, GCs, dSphs) 
seem to suggest/hint that at least two classes of objects contributed to the
building up of the Galactic halo. On one hand, some fragments were similar to
the present-day dwarf spheroidal still orbiting our Galaxy, in particular in the
low metallicity regime (see also Tolstoy et al. 2009). On the other hand, a more
important fraction of the halo seems to have formed in larger fragments, whose
higher mass allowed to reach quickly the same metallicity level of the component
of the Galaxy undergoing dissipative collapse at the same epoch. The ensuing
location of first generation stars lost at early epochs in GCs and of the field
stars of the dissipative component is then undistinguishable, at present, and
corroborate this dual channel for the origin of part of the halo.

\begin{acknowledgements}
This work was partially funded by the PRIN INAF 2011 grant ``Multiple
populations in globular clusters: their role in the Galaxy assembly" (PI E.
Carretta). We thank Angela Bragaglia for valuable suggestions and discussions.
We also wish to thank the referee for useful comments.
This research has made extensive use of the NASA's Astrophysical Data System and
the SIMBAD database (in particular Vizier), operated at CDS, Strasbourg, 
France, without which this work could not have been done. We think that these 
databases are fundamental tools for advance in Astrophysics, and we wish to
warmly thank all colleagues who in true scientific spirit contribute their data
to these facilities.
\end{acknowledgements}

\end{document}